\documentclass[aps,prb,twocolumn,floatfix,groupeaddress]{revtex4}
\usepackage[dvips]{graphicx}
\usepackage{amsmath}
\usepackage{dcolumn}
\usepackage{bm}

\newcommand{\bra}[1]{\left<#1\right\vert}
\newcommand{\ket}[1]{\left\vert#1\right>}
\newcommand{\braket}[2]{\left<#1\vert#2\right>}
\begin{document}
%\title{Magic, Mystery, and More in Relativistic Electron Energy Loss Spectra}
\title{The Magic Angle ``Mystery" in Electron Energy Loss Spectra: Relativistic
and Dielectric Corrections}
\author{A. P. Sorini}
\affiliation{Department of Physics, University of Washington, Seattle, WA 98195}
\author{J. J. Rehr }
\affiliation{Department of Physics, University of Washington, Seattle, WA 98195}
\author{Z. H. Levine}
\affiliation{National Institute of Standards and Technology, Gaithersburg, Maryland 20899}

\begin{abstract}
Recently it has been demonstrated that a careful treatment of both
longitudinal and transverse matrix elements in electron energy loss spectra
can explain the mystery of relativistic effects on the
{\it magic angle}. Here we show that there is an additional
correction of order $(Z\alpha)^2$ where $Z$ is the atomic number and
$\alpha$ the fine structure constant, which is not necessarily 
small for heavy elements. Moreover, we suggest that macroscopic
electrodynamic effects can give further corrections which can
break the sample-independence of the magic angle.
\end{abstract}
\date{\today}
\maketitle
\section{Introduction}

The title of this article is in reference to a recent work by
Jouffrey {\it et al}.~\cite{jouffrey04} with the title
``The Magic Angle: A Solved Mystery."
The {\it magic angle} in electron energy loss spectroscopy (EELS)
is a special value of the microscope collection-angle $\alpha_c$ 
%in an electron energy-loss (EELS) experiment
at which the measured spectrum ``magically" becomes independent 
%of the sample tilt-angle, even for anisotropic samples.
of the angle between the incoming beam and the sample ``$c$-axis."
%, even for anisotropic samples.
The mystery, in the context of 200 keV electron microscopy, is that
standard semi-relativistic quantum theory 
yields a ratio of the magic angle $\theta_M$ to 
``characteristic angle" $\theta_E$ of more than twice the 
observed~\cite{daniels03} value. Unfortunately, 
time~\cite{paxton}
and again,\cite{daniels03,schatti}
the theoretical justification of the
factor of two turned out to be an errant factor of two elsewhere
in the calculation.
%followed by an application of the well-known result $2/2=1$. 
A key contribution of Jouffrey {\it et al}.~was the observation
that relativistic ``transverse" effects, when properly included in the
theory, naturally give a factor of two correction to the
non-relativistic magic angle.
%Here we show that moreover, there are yet further corrections
Here we show that there are yet additional corrections
to the theory which can even break the sample independence of the
magic angle. 

As in Ref.\ [\onlinecite{jouffrey04}],
%In this paper
we consider here the problem of a relativistic probe
electron scattering off of a macroscopic condensed matter sample.
Similar problems have been solved long ago using both
semi-classical~\cite{moller32} and fully quantum-mechanical
approachs.\cite{bethe30, fano56, fano63}
%etc] 
Indeed, the fully quantum-mechanical, relativistic case 
of scattering two plane-wave electrons has long been 
a textbook problem.\cite{heitler54, peskin95}
This
%Told dead horse
classic problem was revived recently in the works of
%Ttwo interesting papers
Jouffrey {\it et al}.~\cite{jouffrey04}
and of Schattschneider {\it et al}.,\cite{schatt05}
in which a ``flaw" in the standard theory is pointed out. 
The flaw is the approximation that
the so-called ``longitudinal" and ``transverse" matrix elements
for the scattering process may be summed incoherently, as argued by
Fano in a seminal paper.\cite{fano56}
%for an atomic system that is rotationally invarient
 In fact, this approximation is only valid
when the sample under consideration posseses certain symmetries.
In a later review article,\cite{fano63}
Fano states this condition explicitly; namely that his original
formula for the cross-section is only applicable to systems of 
cubic symmetry. However, this caveat,
%which is only mentioned clearly in Fano's later review article,
seems to have been generally ignored, and hence
%and indeed turns out to be the source of 
%and its neglect
turns out to be the source of 
%have been the source of 
the magic angle ``mystery".\cite{jouffrey04}
Jouffrey {\it et al.}, and later
Schattschneider {\it et al}.,~showed that if one correctly sums and squares
the transition matrix elements then, in the dipole approximation,
one finds the magic angle corrected by a factor of two.

%! word "aim" repeated
%Our aim here is to examine the theory in more detail with the aim
Our aim here is to examine the theory in more detail in order to
%of deriving the relativistic corrections to the magic angle.
derive both relativistic and material-dependent corrections to the magic angle.
In Section II we consider relativistic electron scattering within the
formalism of 
%Coulomb gauge 
quantum electrodynamics (QED). Working the Coulomb gauge, 
we show that one can almost reproduce the results of Jouffrey {\it et al}.~and
the theory of Schattschneider {\it et al}., apart from a simple correction 
%term of order ${\omega}/{mc^2}$, which is not always negligible.
term of order ${\hbar\omega}/{mc^2}$, which is not always negligible. Here
$\hbar\omega$ is the energy lost by the probe and $mc^2$ is the 
rest energy of an electron.
In Section III we suggest the possibility of incorporating macroscopic
electrodynamic effects into the theory, which can break the symmetry
of sample independence of the magic angle.

%one may use a so-called 
%"relativistically corrected schrodinger equation" [pdw]
%instead of the Dirac or Klein-Gordan equations to treat the fast probe
%electron. This particular approximation works well for electrons of energies
%less than about 500keV [fujiwara]. This means that, if we please, we may
%treat both the sample and the probe electrons as each obeying
%a schrodinger equation where the latter schrodinger equation is actually 
%"relativistically corrected." In this note we will treat the sample in the
%independent particle approximation. We use these approximations only in order
%to simplify our notation and we will intermitently remark on how to 
%generalize our results beyond these approximations.
%
%%obeys its own Dirac (or appropriate Schrodinger) equation. What this means
%in terms of "diagrams" is that electron propagators are never included in
%loops or integrated over. In fact, we shall only see electrons as external
%lines. This is perfectly adequate for our purposes. In fact, there should
%be no reason for us to ever quantize our probe field. It would, on the
%other hand, be interesting to included the many-body sample effects 
%by quantizing the sample electron field. This issue will not be 
%

\section{Coulomb Gauge Calculation}

An appealing aspect of the formalism of Schattschneider {\it et al}.~is 
its
%surprising
simplicity. Their approach is similar to
the semi-classical approach of M\o ller,\cite{moller32} but with 
the added simplification of working with a probe
and sample described by the Schr\"odinger equation, rather than the Dirac 
equation. They also find that the theory is simplified by choosing to
work in the Lorentz gauge.  Unfortunately, however, the theory of M\o ller
is somewhat {\it ad hoc} in that a classical calculation in the 
Lorentz gauge
is modified by replacing the product of two classical charge
densities by the product of four different wavefunctions in order
to obtain the transition matrix element.  For the M\o ller case
this procedure is justified {\it a posteriori}
by the fact that it reproduces the
correct result, but is only rigorously justified by appealing
to the method of second quantization.\cite{heitler54}
%
%We feel that
M\o ller's proceedure is physically reasonable {\it a priori},
because M\o ller was interested in the scattering of electrons in
vacuum. However, the theory of Schattschneider {\it et al.,}~which
largely mimics M\o ller's theory, is less physically reasonable {\it a priori},
since the electrons are not scattering in vacuum, but are
inside a solid which can screen the electrons.  Nevertheless, since
the discrepancy is small, the Schattschneider {\it et al.,} theory
is justified {\it a posteriori} to a lesser extent by
experiment.\cite{daniels03}  We thus refer to the theory of Schattschneider et
{\it al}. as a ``vacuum-relativisitic theory." Consequently, in an effort to
account for the discrepancy with experiment, we feel that it is useful
to rederive the results of Jouffrey {\it et al}.\ from a more fundamental
starting point.
%and see if anything interesting turns up.

%As it turns out, the theory of Schattschneider {\it et al}.~is not exact,
It is easy to see that the theory of 
Schattschneider {\it et al}.~is not formally exact,
though for many materials the error in the vacuum relativistic limit
is negligible. In fact, the discrepancy 
%slightly incorrect, but only in such a way that
can be easily explained via single-particle
quantum mechanics: although Schattschneider
{\it et al}.~work explicitly in the Lorentz gauge, they also 
make the assumption that the momentum and the vector potential
commute,
\begin{equation}
%{\bm p}\cdot{\bm A({\bm r})}={\bm A({\bm r})}\cdot{\bm p}.
{\bm p}\cdot{\bm A({\bm r})}\stackrel{?}{=}{\bm A({\bm r})}\cdot{\bm p}.
\end{equation}
Of course, this commutation relation is only exact in the Coulomb gauge.
In the end, however, the error in this approximation
only effects the final results (e.g., matrix elements) by a correction
of order $\hbar\omega/mc^2$ compared to unity, where $\hbar\omega$ is the 
energy lost by the probe.  Since $\hbar\omega/mc^2$ is at most
$(Z\alpha)^2$ for deep-core energy loss, the effect is
usually negligible, except of course, for very heavy atoms. To see % exactly
how corrections such as the above enter into the theory, and 
further to determine 
whether or not such corrections are meaningful
or simply artifacts of the various approximations used
in the theory of Schattschneider et {\it al}.,
we find it useful to present a fully quantum-mechanical,
relativistic many-body treatment along the lines of Fano,\cite{fano63}
but without any assumption of symmetry of the sample. 
Our treatment is at least as general as that of Schattschneider
{\it et al}.~as far as the symmetry of the sample is concerned. Thus
going beyond the formulations of Schattschneider {\it et al}.~and M\o ller,
we take as our starting point the many-particle QED Hamiltonian.
We then show that in a single-particle approximation
the theory yields the result of
% rather than a semi-classical starting point.
Schattschneider {\it et al}.~together with the correction mentioned above.

Our starting point therefore is the Hamiltonian
in Coulomb gauge~\cite{heitler54}
\begin{equation}\label{Htot}
H=H_{\textrm{el}}+H_{\textrm{int}}+H_{\textrm{rad}}\;,
\end{equation}
where the Hamiltonian has been split into three parts:
\noindent
i) the unperturbed electron part 
\begin{equation}\label{Eq:Hel}
H_{\textrm{el}}=\int d^3x\,\psi^{\dagger}({\bm x})
\left(
c{\bm \alpha} \cdot{\bm p}+{\bm \beta} mc^2
\right)
\psi({\bm x}),
\end{equation}
where $\psi({\bm x})$ is the second-quantized Dirac field, ${\bm \alpha_i}$ and 
${\bm \beta}$ are the usual Dirac matrices, $m$ is the electron mass, and 
$c$ is the speed of light;

\noindent
ii) the unperturbed (transverse) radiation part 
\begin{equation}\label{Eq:Hrad}
H_{\textrm{rad}}=\sum_{{\bm k}}\sum_{i=1}^2a^{\dagger}_{{\bm
k},i}a_{{\bm k},i}\hbar\omega_k,
\end{equation}
where $a_{{\bm k},i}$ destroys a photon of momentum ${\bm k}$, 
polarization ${\bm \epsilon}_{{\bm k},i}$, 
and energy $\hbar \omega_k$; and

\noindent
iii) the interaction part
\begin{eqnarray}\label{Eq:Hint}
H_{\textrm{int}}&=&
+\ e \int d^3x\,\psi^{\dagger}({\bm x}){\bm \alpha}\cdot{\bm A({\bm x})}
\psi({\bm x}),
\nonumber
\\
&+&\frac{e^2}{2}\int d^3x d^3y 
\frac{\psi^{\dagger}({\bm x})\psi^{\dagger}({\bm y})
\psi({\bm y})\psi({\bm x})}
{|{\bm x}-{\bm y}|},
\end{eqnarray}
where
\begin{equation}\label{Avec}
{\bm A}({\bm x})=\sum_{{\bm k},{i}}
  \sqrt{\frac{2\pi\hbar c^2}{V\omega_k}}
\left(
a_{{\bm k},{i}}{\bm \epsilon}_{{\bm k},i}
e^{i{\bm k}\cdot{\bm x}}
+
a_{{\bm k},i}^{\dagger}{{{\bm \epsilon}}^*}_{{\bm k},i}
e^{-i{\bm k}\cdot{\bm x}}
\right),
\end{equation}
$e=|e|$ is the charge of the proton, and $V$ is the system volume.
%of volume $V$.

Let us next specialize to the case of a fixed number ($N+1$) of 
electrons where the ($N+1$)-th electron is singled out as 
the ``fast probe" traveling with velocity ${\bm v_0}$, and the remaining
$N$ electrons make up the sample.
We also introduce a lattice or cluster
of ion-cores (below we consider only elemental solids of atomic number $Z$ but
the generalization to more complex systems is obvious) 
which is treated classically, and which gives rise to a
potential $v_{\textrm{e-core}}({\bm x})=
\sum_{i=1}^{N/Z}(-Ze^2)/|{\bm x}-{\bm R_i}|$ as seen by the electrons. 
In this case our Hamiltonian becomes:
\begin{eqnarray}\label{hammy}
H&=&
\left[
c{\bm \alpha}\cdot({\bm p}+\frac{e}{c}{\bm A}({\bm r}))+{\bm \beta} mc^2
\right] + v_{\textrm{e-core}}({\bm r})
\nonumber
\\
&+&
\sum_{i=1}^N
\left[
c{\bm \alpha^{(i)}}\cdot({\bm p^{(i)}}+\frac{e}{c}{\bm A}({\bm r^{(i)}}))
+{\bm \beta^{(i)}} mc^2
\right]
%+ v_{\textrm{e-core}}({\bm r})
\nonumber
\\
&+&
e^2\sum_{i=1}^N\frac{1}{|{\bm r}-{\bm r^{(i)}}|}
+
\frac{e^2}{2}\sum_{1=i\neq j=1}^N\frac{1}{|{\bm r^{(i)}}-{\bm r^{(j)}}|}
\nonumber
\\
&+&
\sum_{i=1}^N v_{\textrm{e-core}}({\bm r^{(i)}})
+ v_{\textrm{core-core}}+H_{\textrm{rad}}\;,
\end{eqnarray}
where the coordinates which are not labelled by an index 
refer to the probe electron. 
The interaction $v_{\textrm{core-core}}$ between 
ion cores is a constant and is henceforth
dropped.
%the coulomb interactoin is actually contained in teh longitudinal part 
%of the radiation hamiltonian. thus the remaining "free radiation" 
%hamiltonian is actually the transverse part.]
%We are in $\nabla\cdot{\bm A}=0$ gauge and so the interaction is just
%the instantanious coulomb interaction and the transverse photon interaction 
%which couples to the electrons via the term $A$ where
%$$
%{\vec A}({\bm x})=\sum_{{\bm k}\alpha}\sqrt{\frac{2\pi\hbar c^2}{V\omega_k}}
%\left(
%a_{{\bm k}\alpha}{\vec \epsilon}_{{\bm k}\alpha}
%e^{i{\bm k}\cdot{\bm x}}
%+
%a_{{\bm k}\alpha}^{\dagger}{\vec \epsilon}_{{\bm k}\alpha}^{*}
%e^{-i{\bm k}\cdot{\bm x}}
%\right)
%$$
%and the operators $a^{\dagger}$ create free photons.

%The ion
%cores are treated classically and so we drop that term of the hamiltonian as
%it is just a constant. 

To proceed to a single-particle approximation for the sample, 
the interaction of the sample electrons among themselves and with the 
potential of the ion cores may be taken into account 
by introducing a single-particle self-consistent potential $v({\bm x})$ which 
includes both $v_{\textrm{e-core}}({\bm x})$ and exchange-correlation
effects. 
The interaction of the
probe electron with the effective single electron of the sample 
will be considered explicitly. The difference between this interaction
and the actual interaction between the probe and sample can be accounted for
by introducing another potential $v'({\bm x})$ which is not necessarily
the same as $v({\bm x})$; $v'({\bm x})$ is, in theory, ``closer" to the 
pure $v_{\textrm{e-core}}({\bm x})$ potential than $v({\bm x})$ though, in
practice, this difference may not be of interest
(see the Appendix for further explanation of this point). 
The potential $v'({\bm x})$ leads to diffraction of the probe electron,
which will not be considered in this paper in order to make
contact with the theory of Schattschneider {\it et al}.~It is also for this 
reason that we have introduced a single-particle picture of the 
sample, along with the fact that we want to apply this
theory to real condensed matter systems in a practical way. The
extension to the many-body
case, in which the only single-body potential seen by the
probe is due to the ion-cores, is given in the Appendix.
Thus using the single-particle approximation for the sample,
\begin{eqnarray}\label{Eq:Hsing}
H&=&
\left[
c{\bm \alpha}\cdot[{\bm p}+\frac{e}{c}{\bm A}({\bm r})]+{\bm \beta} mc^2
\right]
\nonumber
\\
&+&
\left[
c{\bm \alpha_s}\cdot[{\bm p_s}+\frac{e}{c}{\bm A}({\bm r_s})]
+{\bm \beta_s} mc^2
\right]
\nonumber
\\
&
+& v'({\bm r})
+
e^2\frac{1}{|{\bm r}-{\bm r_s}|}
+
v({\bm r_s})
+
H_{\rm rad},
\end{eqnarray}
where the quantities labeled by the letter $s$ refer to the sample electron
and the unlabeled quantites refer to the probe electron.
%As stated above, 
In the remainder of this paper we set 
$v'\to 0$, though the generalization of the theory to include diffraction
is not expected to be difficult. 

%In our perturbation theory the unperturbed states are then direct products of
%the unperturbed sample electron states (as calulated by feff or something) and 
%the unperturbed probe electron states (plane wave, ignoring diffraction) and 
%the free photon states.

As it turns out,\cite{fujiwara61}  we may start from an effective Schr\"odinger 
treatment of both the sample {\em and the probe} rather than a Dirac 
treatment. The treatment of the probe by a 
``relativistically corrected"
Schr\"odinger equation is standard practice~\cite{pdw04} in much of EELS 
theory, and is appropriate~\cite{fujiwara61} for modern microscope
energies of interest here ({\it e.g.,}  a few hundred keV).
The relativistic correction to the Schr\"odinger equation of the probe
consists in simply replacing the
mass of the probe $m$ by the relativistic mass $m'=m\gamma$
where $\gamma=1/\sqrt{1-v_0^2/c^2}$.  Moreover working with 
a Schr\"odinger equation treatment facilitates
contact with the ``vacuum-relativistic" magic-angle theory of
Schattschneider {\it et al}. We will indicate later how
the results change if we retain a full
Dirac treatment of the electrons. Thus we may start
with the Hamiltonian
\begin{eqnarray}\label{Eq:Hscho}
H
&=&
\frac{[{\bm p}+({e}/{c}){\bm A}({\bm r})]^2}{2m'}
+\frac{[{\bm p_s}+({e}/{c}){\bm A}({\bm r_s})]^2}{2m}
\nonumber
\\
&+&v({\bm r_s})+\frac{e^2}{|{\bm r_s}-{\bm r}|}+H_{\textrm{rad}}
\nonumber
\\
&\equiv&H_0 + \frac{e}{m'c}{\bm p}\cdot {\bm A}({\bm r})+\frac{e}{mc}{\bm p_s}\cdot 
{\bm A}({\bm r_s})
\nonumber
\\
&+&\frac{e^2}{|{\bm r_s}-{\bm r_p}|}+O(A^2)\;.
\end{eqnarray}

In this theory the unperturbed states are then direct products of
unperturbed sample electron states (which in calculations can be 
described, for example, 
by the computer code \textsc{FEFF8},\cite{ankudinov98})
unperturbed probe electron states (plane-waves, ignoring diffraction), 
and the free (transverse) photon states. Also, from now on we ignore the 
interaction terms which are $O(A^2)$. Thus our perturbation is
\begin{equation}\label{Eq:pert}
U=\frac{e^2}{|{\bm r}-{\bm r_s}|}+\frac{e}{m'c}\, {\bm p} \cdot {\bm A}({\bm r})
+\frac{e}{mc}\, {\bm p_s}\cdot {\bm A}({\bm r_s}),
\end{equation}
and we are interested in matrix elements of
\begin{equation}\label{Eq:T}
U+UG_0U+\ldots
\end{equation}
where the one-particle Green's function is
\begin{equation}\label{green}
G_0(E)=\frac{1}{E-H_0+i\eta}
\end{equation}
and $\eta$ is a postive infinitesimal.
The matrix elements are taken between initial and final states 
(ordered as: probe, sample, photon) 
\begin{equation}\label{kets}
|I\rangle = \ket{k_I}\ket{i}\ket{0} \quad {\rm and}  \quad
|F\rangle = \ket{k_F}\ket{f}\ket{0}\;.
\end{equation}
\noindent
To lowest order ($e^2$) there will be a ``longitudinal" (instantaneous Coulomb) 
contribution to the matrix element, and a ``transverse" (photon mediated)
contribution, as illustrated in Fig.~1.
%%for diagrams of these processes.
%\begin{figure}
%\begin{center}
%\includegraphics[scale=0.7]{dgrams.eps}
%%\includegraphics[scale=0.35]{diagram.eps}
%\caption{\label{regions}The lines labelled by momenta $k_I$ and $k_F$ represent the probe particle. The lines labelled by letters $i$ and $f$ represent the sample particle. The dashed line is the instantaneous Coulomb interaction. The wiggly lines are photons. Time flows to the right.}
%\end{center}
%\end{figure}
\begin{figure}
\includegraphics[scale=0.6]{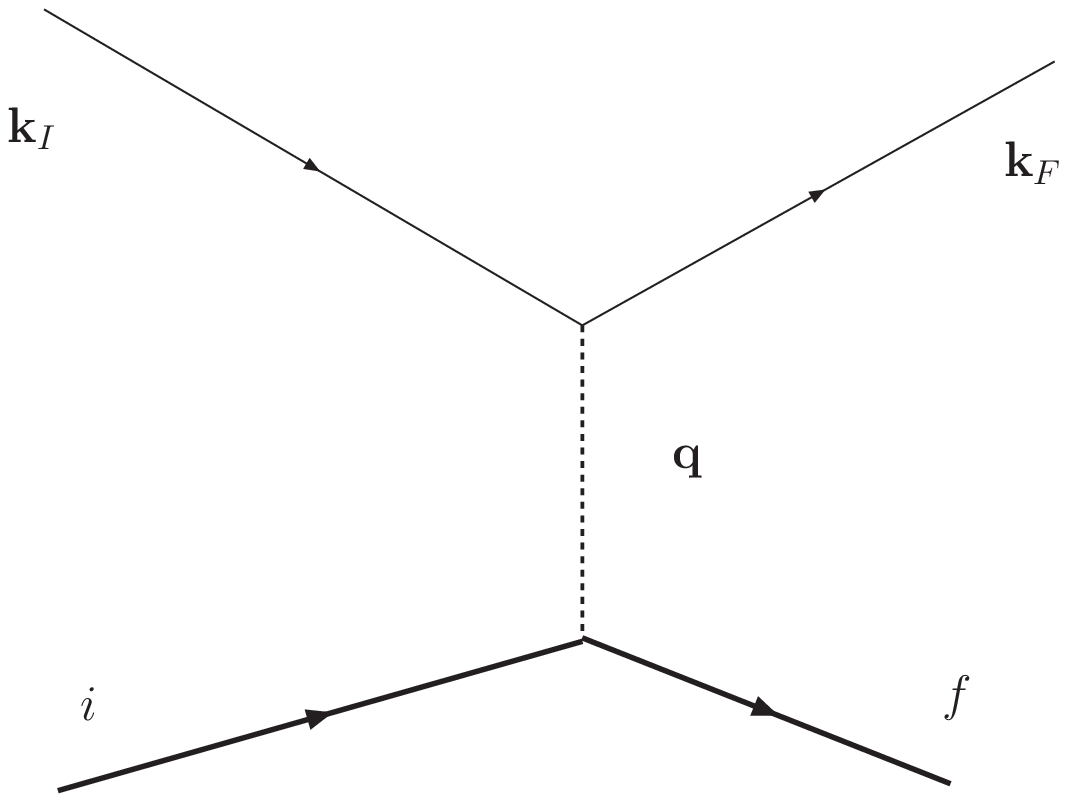}
\includegraphics[scale=0.6]{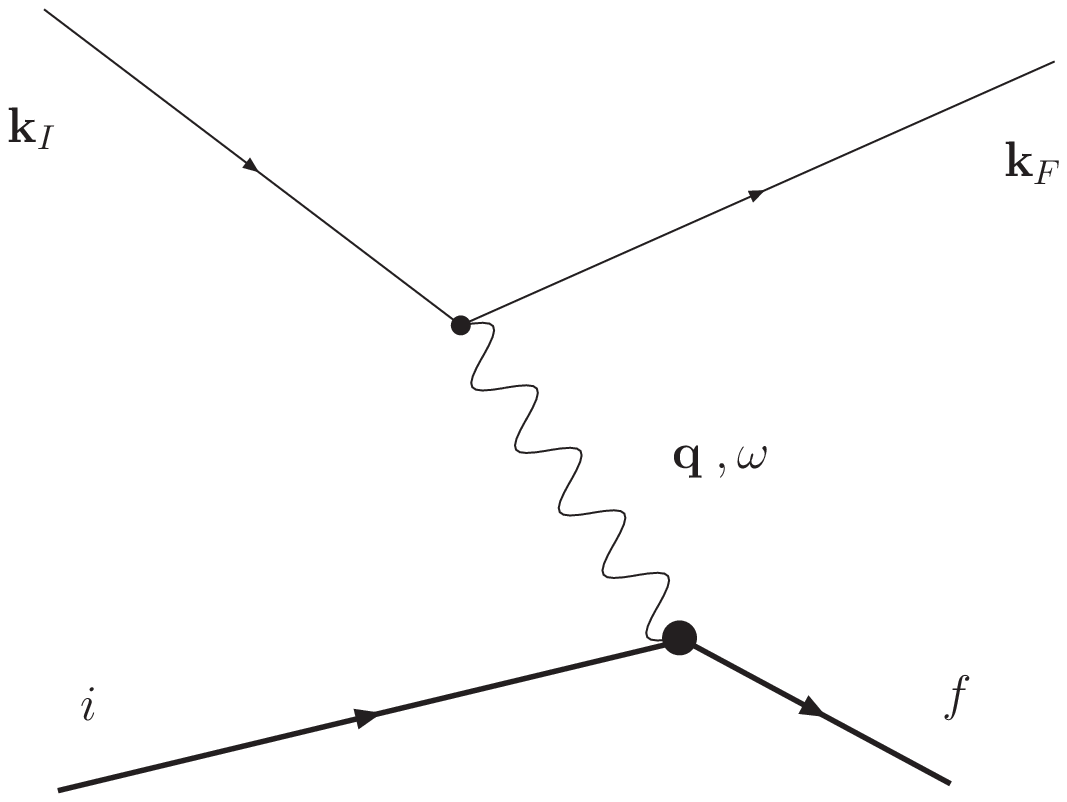}
\includegraphics[scale=0.6]{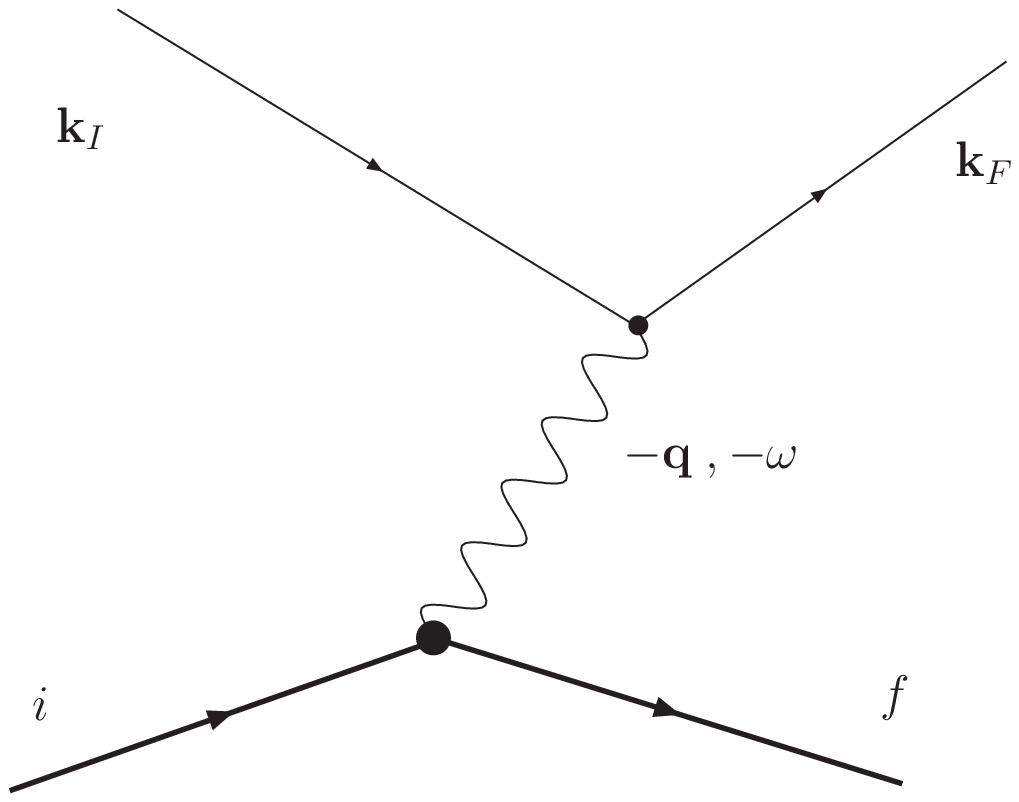}
\caption{\label{regions}Feynman diagrams for the scattering process 
due to both the instantaneous Coulomb interaction (upper) and 
the transverse photon interaction (middle, lower). The solid lines labeled 
by momenta $\textbf{k}_I$ and $\textbf{k}_F$ represent the probe particle; 
thick solid lines labeled by the letters $i$ and $f$ represent the 
sample particle; the dashed line is the instantaneous Coulomb interaction;
and the wiggly lines are transverse photons. Time flows to the right.}
\end{figure}

Instead of elaborating
%of going through all the tedious
the details from standard perturbation theory,
we simply write down the result for the matrix element
%I can just explain what each line in the diagram means (the meaning
%of which are obtained, of course, by going through the tedious details):
%
%The dashed line (which I forgot to label--it only needs 
%a momentum label $\vec p$)
%contributes as
%$$
%\frac{4\pi e^2}{V p^2}
%$$
%and the vertex (and external lines) of the 
%sample electron from $\ket{i}$ to $\ket{f}$ contributes
%as 
%$$
%\bra{f}e^{i{\bm q}\cdot {\bm r_s}}\ket{i}\;;
%$$
%The wiggly line labelled by momentum $\vec p$ and energy $\omega$ 
%contributes
%$$
%\frac{1}{\omega - c |\vec p|}\left(\delta_{ij}-\frac{p_ip_j}{p^2}\right)\frac{2 \pi \hbar c}{V|\vec p|}
%$$
%and connects vertices between probe electron states of momenta $\vec k_i$
%and $\vec k_f$ as
%$$
%k_f^i\frac{e}{mc}
%$$
%(n.b., whether $k_f$ or $k_i$ appears in the above expression does not matter)
%and vertices between sample electron states $\ket{i}$ and $\ket{f}$ as
%$$
%\bra{f}p_s^je^{i{\bm q}\cdot {\bm r_s}}\ket{i}\frac{e}{mc}\;.
%$$
%
\begin{eqnarray}\label{ME}
M &=&\frac{4\pi e^2}{V}\left[
\frac{1}{q^2}\bra{f}e^{i{\bm q}\cdot {\bm r_s}}\ket{i} \right. \nonumber\\
&+&
\left.\frac{1}{\omega^2-c^2q^2}\frac{k_T^j}{m'}
\bra{f}\frac{p_s^j}{m}e^{i{\bm q}\cdot{\bm r_s}}\ket{i}
\right]\;,
\end{eqnarray}
where ${\bm k_T}$ (see Fig.~2) is the part of the initial
(or final) momentum which is
perpendicular to the momentum transfer $\hbar{\bm q}$. In the 
remainder of this paper 
we will choose our units such that $\hbar=1$.
%n.b. those triangles in Fig.~2 are similar. I.e., k_T/k_I = q_\perp/q
\begin{equation}\label{ks}
k_T^j=\left(\delta_{lj}-\frac{q_lq_j}{q^2}\right)k_F^l
=\left(\delta_{lj}-\frac{q_lq_j}{q^2}\right)k_I^l\;.
\end{equation}
\begin{figure}
\begin{center}
\includegraphics[scale=0.55]{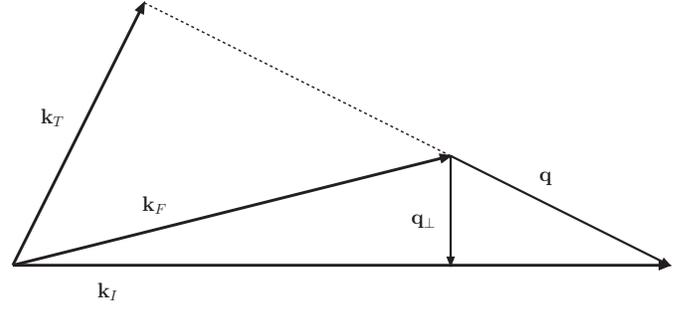}
\caption{\label{transverse}The relevant momenta: ${\bm k_I}$ is the initial momentum of the probe particle, ${\bm k_F}$ is the final momentum of the probe particle, ${\bm q}$ is the momentum transfer ${\bm k_I}-{\bm k_F}$ 
and ${\bm k_T}$ is the part of both the initial of final momenta which is perpendicular to the momentum transfer.}
\end{center}
\end{figure}

The result of Eq.~(\ref{ME})
is easy to understand diagramatically. For example, to each
wiggly line of momentum ${\bm q}$ and energy $\omega$ we may assign a value
\begin{equation}\label{diagex}
\frac{1}{\omega - c |{\bm q}|}
\left(\delta_{ij}-\frac{q_iq_j}{q^2}\right)
\frac{2 \pi c}{V|{\bm q}|} \;.
%\frac{2 \pi \hbar c}{V|{\bm q}|} \;.
\end{equation}
%Et cetera, for other parts of the diagrams.
At this point we note that the relativistic many-body
version of
Eq.~(\ref{ME}) can be obtained by making intuitively 
reasonable replacements such as
${\bm p}/{m} \to c{\bm \alpha }$, 
$e^{i{\bm q}\cdot{\bm r_s}} \to \sum_i e^{i{\bm q}\cdot{\bm r^{(i)}}}$.
See the Appendix for further details. 

Eq.~(\ref{ME}) is equivalent to the matrix elements
given by Fano in Eq.~(12) of Ref.\ [\onlinecite{fano63}]. 
%and is the correct matrix element (atomic, neglecting dielectric).
%does not include any offending ``incoherent" approximation. 
%The offending equation is given as 
The cross-section given by Fano in Eq.~(16) of Ref.\ [\onlinecite{fano63}],
in which the matrix elements have been summed incoherently, 
is not generally correct and is the
source of the magic angle ``mystery".\cite{schatt05}

Before continuing to the dipole approximation it is useful to rewrite 
Eq.~(\ref{ME}) using the definition
\begin{equation}\label{Eq:kT}
{\bm k_T}={\bm k_I}-{\bm q}\frac{{\bm q}\cdot{\bm k_I}}{q^2}
\end{equation}
to eliminate ${\bm k_T}$ in favor of ${\bm k_I}$ 
(or equivalently ${\bm v_0}={\bm k_I}/m'$). 
Making this replacement we obtain
\begin{eqnarray}\label{ME2}
M&=&\frac{4\pi e^2}{V}\left[
\frac{1}{q^2}\bra{f}e^{i{\bm q}\cdot {\bm r}}\ket{i}
-
\frac{{\bm q}\cdot{\bm v_0}}{mq^2}
\frac{\bra{f}{\bm q}\cdot{\bm p} e^{i{\bm q}\cdot{\bm r}}\ket{i}}
{\omega^2-c^2q^2}
\right.
\nonumber
\\
&+&
\left.
\frac{\bra{f}{\bm v_0}\cdot({\bm p}/m)e^{i{\bm q}\cdot{\bm r}}\ket{i}}
{\omega^2-c^2q^2}
\right]\;,
\end{eqnarray}
which can be rewritten as:
\begin{eqnarray}\label{ME3}
M&=&\frac{4\pi e^2}{V}
\frac{1}{q^2-(\omega^2/c^2)}
\bra{f}
e^{i{\bm q}\cdot{\bm r}}
\times
\nonumber
\\
&&\left[
1-\frac{{\bm v_0}\cdot{\bm p}}{mc^2}
-\frac{\omega^2}{q^2c^2}(1-\frac{{\bm q}\cdot{\bm p}}{m\omega})
\right]
\ket{i}\;,
\end{eqnarray}
where we have made use of ${\bm q}\cdot{\bm v_0}=\omega$ in order to
cancel certain terms which appear after commuting the exponential 
through to the far left. Also, we have removed the label $s$ from the
position and momentum of the sample electron. This change in notation
will be used throughout the remainder of this paper.

Eq.~(\ref{ME3}) is the same as
Eq.~(6) of Schattschneider {\it et al}., except for an ``extra" term 
\begin{equation}\label{extra}
\bra{f}
e^{i{\bm q}\cdot{\bm r}}
\left(1-\frac{{\bm q}\cdot{\bm p}}{m\omega}\right)
\ket{i}\;.
\end{equation}
Fortunately, this term may be simplified 
by considering the commutator
\begin{equation}\label{commie}
[e^{i{\bm q}\cdot{\bm r}},H_0]
=[e^{i{\bm q}\cdot{\bm r}},\frac{p^2}{2m}]
=e^{i{\bm q}\cdot{\bm r}}
\left(-\frac{{\bm p}\cdot{\bm q}}{m}-\frac{q^2}{2m}\right)\;,
\end{equation}
where the first equals sign follows from the fact that 
$e^{i{\bm q}\cdot{\bm r}}$ commutes with everything in $H_0$ except for
the kinetic term of the sample electron (by its definition $H_0$ explicitly
contains only local potentials).
%(In the above equation, the $r$ and $p$ refer to the sample and so 
%the $e^{iqr}$ commutes with everything in the unperturbed sample 
%Hamiltonian except for the kinetic term. And I'm using the
%unperturbed sample Hamiltonian because the matrix element has
%been reduced down to one between sample states only... and also
%I will use the condition that $E_i+\omega-E_f=0$ where the $E_i$ and
%$E_f$ are initial and final energies of the sample.)
%$$
%=\frac{-1}{2m}(e^{i{\bm q}\cdot{\bm r}}{\bm p}\cdot{\bm q}
%+{\bm p}\cdot{\bm q}e^{i{\bm q}\cdot{\bm r}})
%=e^{i{\bm q}\cdot{\bm r}}(\frac{-1}{m}{\bm p}\cdot{\bm q}-\frac{q^2}{2m})
%$$
Then, using the fact that for any operator $O$,
\begin{equation}\label{ocom}
\bra{f} [O,H_0]
\ket{i}
=\bra{f}O
\ket{i}(E_i-E_f)
=\bra{f} O
\ket{i}(-\omega)
\end{equation}
we have
\begin{equation}\label{Eq:com}
\bra{f}
e^{i{\bm q}\cdot{\bm r}}
\ket{i}(-\omega)
=
-\bra{f}
e^{i{\bm q}\cdot{\bm r}}
(\frac{{\bm p}\cdot{\bm q}}{m}+\frac{q^2}{2m})
\ket{i}
\end{equation}
and thus 
%(dividing both sides by $-\omega$ and bringing the p.q term to the
%LHS)
\begin{equation}\label{Eq:com2}
\bra{f}
e^{i{\bm q}\cdot{\bm r}}
(1-\frac{{\bm p}\cdot{\bm q}}{m\omega})
\ket{i}
=
\bra{f}
e^{i{\bm q}\cdot{\bm r}}\frac{q^2}{2m\omega}
\ket{i}\;.
\end{equation}
Making the above replacement in Eq.~(\ref{ME3})
we find
\begin{eqnarray}
M&=&\frac{4\pi e^2}{V}
\frac{1}{q^2-\omega^2/c^2} \nonumber \\
 &\times& \bra{f}
e^{i{\bm q}\cdot{\bm r}}
\left(
1-\frac{{\bm v_0}\cdot{\bm p}}{mc^2}
-\frac{\omega^2}{q^2c^2}\frac{q^2}{2m\omega}
\right)
\ket{i}
\end{eqnarray}
%the $q^2$ in that last term cancel and so does one $\omega$ and 
and we see that the 
``extra" term only changes the result by order 
$\omega/mc^2$ where $mc^2$ is the rest energy of an electron and
$\omega$ is the energy lost; 
%and the matrix element of Eq.~(\ref{ME3}) is equal to:
\begin{eqnarray}\label{MEE}
M=
%&=&
\frac{4\pi e^2}{V}
\frac{1}{q^2-\omega^2/c^2}
\bra{f}
e^{i{\bm q}\cdot{\bm r}}
\left(
1-\frac{{\bm v_0}\cdot{\bm p}}{mc^2}
-\frac{\omega}{2mc^2}
\right)
\ket{i}
&&
\nonumber
\\
%&=&
=
\frac{4\pi e^2}{V}
\frac{1}{q^2-\omega^2/c^2}
\bra{f}
e^{i{\bm q}\cdot{\bm r}}
\left(
1-\frac{\bm v_0}{mc^2}\cdot({\bm p}+\frac{{\bm q}}{2})
\right)
\ket{i}.\;\;&&\;\;
\end{eqnarray}
%(the last equals sign uses $\omega={\bm q}\cdot{\bm v_0}$)
Eq.~(\ref{MEE}) is the same as 
what Schattschneider {\it et al}.~would have obtained
if they had not neglected the commutator $[{\bm p},{\bm A}]$.

%
% I added a bit more motivation for why the p+q/2 appears and 
% explaination for why it should appear the way it does. maybe this
% is a bit of overkill as far as explaining goes...
%
That a term proportional to ${\bm p}+{\bm q}/2$ rather than simply
${\bm p}$ appears in Eq.~(\ref{MEE}) is correct and
can be understood from the following simple example: 
The interaction Hamiltonian for a point particle with an
external field is given by $e\phi-e{\bm v}\cdot{\bm A}/c$, or rather
\begin{equation}\label{Eq:intsim}
H_{\rm int}\sim
\int d^3x 
\left(
n({\bm x})\phi({\bm x}) - \frac{1}{c}{\bm j}({\bm x})\cdot{\bm A}({\bm x})
\right)\;,
\end{equation}
where $n({\bm x})$ is the density and ${\bm j}(\bm {x})$ is the current, and where the above integral, with the potentials considered as functions of
the source location, is a convolution in space and thus a product in 
Fourier space--the rough correspondence indicated by the ``$\sim$" symbol
in Eq.~(\ref{Eq:intsim}) is considered more rigourously in the Appendix. 
Next, we note that the Fourier transform of the current density (in 
second-quantization) is given for a free particle by~\cite{mahan81}
\begin{equation}
{\bm j} ({\bm q})=\frac{1}{mV}\sum_{\bm k}
\left(
{\bm k}+\frac{\bm q}{2}
\right)
c^{\dagger}_{{\bm k+q}}c_{\bm k}\;,
\end{equation}
where
\begin{equation}
\psi({\bm x})=\sum_{{\bm k}}c_{\bm k} e^{i{\bm k}\cdot{\bm x}}\;,
\end{equation}
and where {\bf q} is considered to be the momentum transferred {\em to} the
sample. This is in agreement with the usual conventions of EELS
\begin{equation}
{\bm q}={\bm k_I}-{\bm k_F}\;.
\end{equation}
Thus we see that Eq.~(\ref{MEE}) is indeed correct, in both sign and 
magnitude of the ``extra" term.

%
% the q/2 can be understood by Fourier transforming the
% j.A part of the interaction and remembering the correct 
% form of j
%

%Indeed, these cancelations are amusing in
%that the exponential may be placed on either side of the other
%operator in the matrix element:
%\begin{eqnarray}
%M&=&\frac{4\pi e^2}{V}
%\frac{1}{q^2-\omega^2/c^2}
%\bra{f}
%e^{i{\bm q}\cdot{\bm r}}
%\left(
%1-\frac{{\bm v_0}\cdot{\bm p}}{mc^2}
%-\frac{\omega^2}{q^2c^2}(1-\frac{{\bm q}\cdot{\bm p}}{m\omega})
%\right)
%\ket{i}
%\nonumber
%\\
%&=&
%\frac{4\pi e^2}{V}
%\frac{1}{q^2-\omega^2/c^2}
%\bra{f}
%\left(
%1-\frac{{\bm v_0}\cdot{\bm p}}{mc^2}
%-\frac{\omega^2}{q^2c^2}(1-\frac{{\bm q}\cdot{\bm p}}{m\omega})
%\right)
%e^{i{\bm q}\cdot{\bm r}}
%\ket{i}\;.
%\nonumber
%\end{eqnarray}

\subsection{Dipole Approximation and the Magic Angle}

In the dipole approximation Eq.~(\ref{MEE}) reduces to
\begin{equation}
\frac{4\pi e^2}{V}\frac{1}{q^2-\omega^2/c^2}
\bra{f}
\left(i{\bm q}\cdot{\bm r}
%-\frac{{\bm v_0}\cdot{\bm p}}{mc^2}
-\frac{{\bm v_0}\cdot{\bm p}}{mc^2}
%i{\bm q}\cdot{\bm r}
\right)
\ket{i}
\;.
\end{equation}
The term $(1-{\bm v_0}\cdot{\bm q}/2mc^2)$ does not contribute because 
$\braket{i}{f}=0$.
Now, we make use of the
replacement $ {\bm p}/{m} \to {i\omega{\bm r}}$ 
which is appropriate within the matrix element to find
\begin{equation}\label{dip}
\frac{4\pi e^2}{V}\frac{i}{q^2-\omega^2/c^2}
\bra{f}
\left({\bm q}-\frac{{\bm v_0}({\bm q}\cdot{\bm v_0})}{c^2}\right)\cdot{\bm r}
\ket{i}\;.
\end{equation}
We have thus found the same ``shortened $q$-vector" that appears in
Eq.~(15) of Schattschneider {\it et al}.~and Eq.~(2) of 
Jouffrey {\it et al}.~Specifically, for
an initial electron velocity ${\bm v_0}$ in the $z$-direction, we have
found the replacement $q_z\to q_z(1-v_0^2/c^2)$
which in turn leads to a significant correction 
(on the order of 100 percent for
typical electron microscopes) to the magic angle. 

The magic angle $\theta_M$ is defined for materials with a ``$c$-axis"
by the equality of two functions of collection angle $\alpha_c$:
\begin{equation}\label{Eq:F}
F(\alpha_c)\equiv\int_0^{\alpha_c} d\theta\theta
%\frac{\theta^2}{{[\theta^2+\theta_E^2(1-v_0^2/c^2)]}^2}
\frac{\theta^2}{{[\theta^2+\theta_E^2/\gamma^{4}]}^2}\;,
\end{equation}
and
\begin{equation}\label{Eq:G}
%G(\alpha_c)\equiv 2\theta_E^2{(1-v_0^2/c^2)}^2\int_0^{\alpha_c} d\theta\theta
%\frac{1}{{[\theta^2+\theta_E^2(1-v_0^2/c^2)]}^2}\;.
G(\alpha_c)\equiv 2\frac{\theta_E^2}{\gamma^{4}}
\int_0^{\alpha_c} d\theta\theta
\frac{1}{{[\theta^2+\theta_E^2/\gamma^{4}]}^2}\;.
\end{equation}
where $\gamma=(1-v_0^2/c^2)^{-1/2}$,
$\theta_E$ is the so-called ``characteristic angle" given in terms
of the energy-loss $\omega$, the initial probe speed $v_0$, and 
$k_I\theta_E=\omega/v_0$.
Both of the above integrals may easily be evaluated in terms of 
elementary functions, but we leave them in the above-form for comparison with 
the theory of the Section III. Eqs.~(\ref{Eq:F}) and (\ref{Eq:G}) both 
make use of the approximation $\sin(\theta)\approx \theta$. Since typical
scattering angles are on the order of milli-radians this small angle 
approximation is highly accurate.

The expressions for $F(\alpha_c)$ and $G(\alpha_c)$
are easily derived within the framework of the Schattschneider
``vacuum theory"\cite{schatt05} and 
result in a ratio of magic angle to characteristic angle which is independent
of the material which makes up the sample. The factors of $(1-v_0^2/c^2)$ which
appear in Eqs.~(\ref{Eq:F}) and (\ref{Eq:G}) come from including the 
transverse effects (as in Section I) and thus the non-relativistic 
($c\to\infty$) result for the ratio of magic angle to characteristic angle
is independent of transverse effects. The ``transverse" 
correction to the magic angle is on the order of 100 percent. This 
corrected theoretical magic angle is in much better agreement with 
the experimentally observed magic angle, although the experimentally
observed magic angle seems be somewhat larger (on the order of 30 percent)
and sample dependent.\cite{daniels03} 
These further discrepancies between theory and experiment are addressed 
in Section III.

\section{Macroscopic Electrodynamic Effects}

%That the result of Schattschneider et al.~is correct in the dipole 
%approximation and even beyond, to order $\omega/mc^2$, suggests 
%the possiblilty of further improvements, by a simple extension 
%%result with a simple change in
 As discussed above, the result of Schattschneider {\it et al}.~is nearly in
 agreement with that obtained in Section II of this paper in
the vacuum relativistic limit.
%approximation and even beyond, to order $\omega/mc^2$, suggests 
%the possiblilty of further improvements, by a simple extension 
However, because of the residual discrepancy between these results
and experiment
%Because of the simplicity of the
we now consider how 
%the single-particle formalism to 
macroscopic electrodynamic effects can be incorported into the
quantum mechanical single-particle formalism.
%with hope of further reducing 
%the remaining discrepancy between the theoretical and experimental 
%values of the magic angle.
We find that the corrections to the magic angle
which result %from the inclusion of dielectric response will be shown to
can be quite substantial at low energy-loss. However, we are unaware of
any experimental data in this regime with which to compare the
theory.
%The only existing magic-angle data is given for values of energy-loss
%well above the region where dielectric response substantially effects
%the results. 
%The only existing magic-angle data is taken at too 
%high of energy-loss for any dielectric response to effect the results. 
Nevertheless, the inclusion of dielectric response introduces a
sample dependence of the theoretical magic angle which
is consistent with the sign of the observed discrepancy.
%actually there arn't too many observations... this suggests experiemtn..
%standard microscopes (~300 keV) but with low energy-loss (~10 eV) on
% samples with c-axis. the change in the magic angle due to the
% dielectric would be profound is this thoery is correct.
 
Certain condensed matter effects
are already present in the existing formalism via the behavior
of the initial and final single-particle states in the sample,
and in many-electron effects which are neglected in the independent
electron theory.  However, the macroscopic response of the sample can be taken
into account straightforwardly within a dielectric formalism.
This procedure is similar to the well-known
``matching" procedure between atomic calculations and 
macroscopic-dielectric calculations of the stopping 
power.\cite{landau84,salvat05,sorini06}
That is, the fast probe may interact with many 
atoms at once, as long the condition $v_0>>\omega_0 a$ (where $\omega_0$ is a
typical electronic frequency and $a$ a typical length scale) is fulfilled.
%possibly more than may be reasonably taken into account in
%a quantum-mechanical cluster calculation. 
%But, on the other hand, because the probe interacts 
%with many atoms at once 
Under these conditions the sample can be treated using the
electrodynamics of continuous media.\cite{landau84} 

Effects due to the macroscopic response 
of the system can be included
within a formalism that parallels that of Schattschneider
{\it et al}.~simply by choosing the 
``generalized Lorentz gauge"\cite{landau84} for a given
dielectric function $\epsilon(\omega)$, instead
of the Lorentz gauge of the vacuum-relativistic theory.
%[Rant: We loose nothing by this. 
%We don't care if our theory is Lorentz invarient
%for god sake--it's a Hamiltonian formalism--and there is certainly a
%prefered frame! Who cares if the interaction
%part of the Hamiltonian 
%(which has no obligation to be Lorentz invarient
%s it is the zero component of a four-vector) is invarient or not?!
%
% 
% Anyways... it is pretty neat that it is *when there is no matter around*
% ... and, of course, in that case it 
% actually is the negative of the interactiong *lagrangian* and the
% lagrangian density if a lorentz scalar... but we don't really care
% and the physics is much more clear, in my opinion, in the 
% coulomb gauge anyways]
In the generalized Lorentz gauge, most of the formal manipulations 
of Schattschneider {\it et al}.~carry through in the same way, 
except that instead of Eq.~(\ref{ME3}) we end up with
%we do not end up with Eq.~(\ref{ME3}). Rather, we end up with:
\begin{eqnarray}\label{MEEPS}
M&=&
\frac{4\pi e^2}{\epsilon(\omega)V}
\frac{1}{q^2-\epsilon(\omega)\omega^2/c^2} \times
\nonumber \\
&&
\bra{f} e^{i{\bm q}\cdot{\bm r}}
\left[
1-\frac{\epsilon(\omega){\bm v_0}}{mc^2}\cdot({\bm p}+\frac{{\bm q}}{2})
\right]
\ket{i}\;.
\end{eqnarray}
The factors of $\epsilon$
in Eq.\ (\ref{MEEPS}) can be understood physically as due to the 
fact that $c\to c/\sqrt{\epsilon}$ in the medium, and also to the fact that
the sample responds to the electric field ${\bm E}$ rather than the electric
displacement ${\bm D}$. 
Eq.~(\ref{MEEPS}) is derived in the following subsection.
%I will work through the Schattschneider formalism in the 
%generalized Lorentz gauge a bit more explicitly in this next
%subsection.

\subsection{Generalized Lorentz Gauge calculation}

We consider a probe electron which passes through 
a continuous medium characterized by a macroscopic 
frequency-dependent dielectric
constant $\epsilon(\omega)$ and magnetic permeability $\mu= 1$. It is
appropriate to ignore the spatial dispersion of the dielectric constant
at this level of approximation.\cite{cockayne06}
Then Maxwell's equations are
\begin{equation}\label{Eq:Max1}
\nabla\cdot{\bm D}=4\pi \rho_{\textrm{ext}},
\end{equation}
with ${\bm D}=\epsilon {\bm E}$. And
\begin{equation}\label{Eq:Max2}
\nabla\times{\bm B}=\frac{4\pi {\bm j_{\textrm{ext}}}}{c}
+\frac{1}{c}\frac{\partial {\bm D}}{\partial t},
\end{equation}
where the charge/current densities $\rho_{\textrm{ext}}$ 
and ${\bm j_{\textrm{ext}}}$ refer only to the ``external" charge and
current for a
probe electron shooting through the material at velocity ${\bm v_0}$.
The other two Maxwell equations refer only to ${\bm E}$ and 
${\bm B}$, and can be satisfied exactly using the definitions
\begin{equation}\label{Eq:Egauge}
{\bm E}=-\nabla\phi-\frac{1}{c}\frac{\partial {\bm A}}{\partial t}\;,
\end{equation}
and
\begin{equation}\label{Eq:Bgauge}
{\bm B}=\nabla\times{\bm A}\;.
\end{equation}
We next insert Eqs.~(\ref{Eq:Egauge}) and (\ref{Eq:Bgauge}) into
Eqs.~(\ref{Eq:Max1}) and (\ref{Eq:Max2})
%and find, well, the usual
%mess, which can be simplified by choosing 
and choose the generalized Lorentz gauge~\cite{landau84}
\begin{equation}\label{Eq:GLGxt}
\nabla\cdot{\bm A}+\frac{1}{c}
\frac{\partial}{\partial t}\int dt'\,\epsilon(t-t')\phi(t')=0\;.
\end{equation}
%The generalized Lorentz gauge in Fourier space reads
%\begin{equation}\label{Eq:GLGqw}
%{\bm q}\cdot{\bm A}-\frac{\epsilon\omega}{c}\phi=0\;.
%\end{equation}
This gauge choice leads to the momentum space (${\bm q}$,$\omega$) equations
\begin{equation}
\left[
-q^2+\epsilon(\omega)\frac{\omega^2}{c^2}
\right]\phi({\bm q},\omega)=4\pi\frac{\rho_{\textrm{ext}}({\bm q},\omega)}{\epsilon(\omega)},
\end{equation}
and
\begin{equation}
\left[
-q^2+\epsilon(\omega)\frac{\omega^2}{c^2}
\right]{\bm A({\bm q},\omega)}
=4\pi\, \frac { {\bm j_{\textrm{ext}}({\bm q},\omega)}}  {c} .
\end{equation}
%And there really is that overall
%``extra" factor of epsilon in the equation for phi
%(cf. Landau\cite{landau84}, Eq.~(114.7)).
%We now specialize to our case of interest by writing
We now write
$\rho_{\textrm{ext}}({\bm q},\omega)=
(-2\pi e)\delta(\omega-{\bm q}\cdot{\bm v_0})$
and
${\bm j_{\textrm{ext}}}={\bm v_0}\rho_{\textrm{ext}}$
to find explicit expressions for $\phi$ and ${\bm A}$:
\begin{equation}
\epsilon(\omega)\phi({\bm q},\omega)
=
\frac{4\pi(-2\pi e)\delta(\omega-{\bm q}\cdot{\bm v_0})}
{[{\epsilon(\omega)\omega^2}/{c^2}]-q^2},
\end{equation}
and
\begin{equation}
{\bm A}({\bm q},\omega)=\frac{ {\bm v_0} }{c} \epsilon(\omega)
\phi({\bm q},\omega)\;.
\end{equation}
Then, proceeding roughly in analogy with Schattschneider {\it et al}.,~we
have
\begin{eqnarray}
H&=&H_0+\frac{e}{2mc}\left({\bm p}\cdot{\bm A}+{\bm A}\cdot{\bm p}\right)
-e\phi
+O(A^2)
\nonumber
\\
&=&
H_0
+
\frac{e}{2mc}\left(
2{\bm A}\cdot{\bm p}-i\nabla\cdot{\bm A}
\right)
-e\phi
+O(A^2).\;\;\;
\end{eqnarray}
%($\hbar=1$, but it's a good placeholder so I might keep that one 
%$\hbar$ around)
Next, evaluating the perturbation $U\equiv H-H_0$
with ${\bm A}=({{\bm v_0}}/{c})\epsilon(\omega)\phi$, we find
\begin{equation}
U=\frac{e}{mc}
\left(
\phi\frac{\epsilon(\omega)}{c}{\bm v_0}\cdot{\bm p}
-i\epsilon\frac{{\bm v_0}}{2c}\cdot\nabla\phi
\right)
-e\phi\;.
\end{equation}
%
%As you can see, I have not bothered to fill in the argument of the 
%function $\phi$ in the above expression. That's because, in my opinion,
%the way this works in the Schattschneider formalism is a little
%shady. So I should try to explain a bit about exactly what they do.
%
% should roughly be same as moller's replacement of two charge
% densityies with four different wave functions to get transition matrix
% elements
%
%A spqm  procedure that works is as follows:
%1. Evaluate the classical potentials of a single point charge moving 
%along. 2. Pretend like this is an external electromagnetic potential
%that gets introduced into the single-particle schrodinger equation in
%the usual way... except that... 
%3. In order to actually get it to couple to {\em two}
%electrons (probe and sample) 
%we take the classical potential and evaluate its spatial argument
%at ${\bm r-r_s}$ where ${\bm r}$ is the position operator of the probe
%and ${\bm r_s}$ is the position operator of the sample particle and 
%evaluate the time argument at zero. 4. Calculate the usual EELS matrix
%elements in the usual way.
%
%Because of the above, 
%
In calculating the matrix element of $U$ 
it is appropriate to replace 
$\nabla\phi$ by $i{\bm q}\phi$ for the case when the final
states are on the {\em left} in the matrix element. Thus 
%
%This is reasonable, but 
%to see that this is true more rigorously (to see that, e.g., the sign
%is right) one could insert 
%position-space unity operators for both the probe and the sample and 
%then note that because the final states are on the left we get a factor
%of $e^{iqr}$ with $q=k_I-k_f$. But the nabla acting on phi is with respect
%to the {\em sample} (because it was a sample momentum commutator which 
%introduced it). Thus, indeed, we pick up 
%an $i{\bm q}$ and not a $-i{\bm q}$.
%Anyways...
%
%I plug and chug and whatnot (I have some more notes on this too), but at
%the end of the day I end up with a matrix element
\begin{eqnarray}\label{GLGME}
M &\equiv& \bra{f}\bra{k_f} U \ket{k_i}\ket{i} =
%-e\phi({\bm q},\omega=({\bm q}\cdot{\bm v_0}))
-e\phi({\bm q},\omega) \nonumber \\
%\left[
%1-\frac{\epsilon(\omega={\bm q}\cdot{\bm v_0})}{mc^2}{\bm v_0}
%\cdot({\bm p}+\hbar{\bm q}/2)
%\right]
%\ket{i}\;.
%\;\;\;\;\;\;\;\;
&\times& \bra{f} e^{i{\bm q}\cdot{\bm r}} \left[
1-\frac{\epsilon(\omega)}{mc^2}{\bm v_0}
\cdot({\bm p}+{\bm q}/2)
\right]
\ket{i}\;.
\;\;\;\;\;\;\;\;
\end{eqnarray}
Alternatively,  since
\begin{equation}
\phi({\bm q},\omega)=\frac{-4\pi e}{\epsilon(\omega)
\left(q^2-{\omega^2\epsilon(\omega)}/{c}\right)}\;,
\end{equation}
we have
\begin{eqnarray}\label{Eq:GLGM}
M
&=&
\frac{4\pi e^2}{\epsilon(\omega)(q^2-\epsilon(\omega)\omega^2/c^2)}
\bra{f}
e^{i{\bm q}\cdot{\bm r}}
\nonumber
\\
&
\times&
\left[
1-\frac{\epsilon(\omega)}{mc^2}{\bm v_0}\cdot({\bm p}+{\bm q}/2)
\right]
\ket{i}\;.
\end{eqnarray}

In the dipole approximation Eq.~(\ref{Eq:GLGM}) reduces to
\begin{equation}\label{dip2}
\frac{4\pi e^2}{\epsilon(\omega)V}\frac{1}{q^2-\epsilon(\omega)\omega^2/c^2}
\bra{f}
i\left[{\bm q}-\epsilon(\omega)\frac{{\bm v_0}({\bm q}
\cdot{\bm v_0})}{c^2}\right]\cdot{\bm r}
\ket{i}\;,
\end{equation}
where $\epsilon(\omega)$ is the generally complex valued macroscopic 
dielectric constant as which can be calculated, for example,
by the \textsc{FEFFOP}~\cite{prangeetal05} code. Consequently
we find that that instead of the longitudinal $q$-vector replacement
\begin{equation}\label{rep1}
{q_z}\to
q_z(1-\beta^2)
\end{equation}
found by Jouffrey {\it et al}.~and Schattschneider {\it et al.},~we 
obtain the replacement
\begin{equation}\label{rep2}
q_z
\to
q_z[1-\epsilon(\omega)\beta^2],
\end{equation}
which is appropriate for an electron traversing a continuous
dielectric medium.  In the same way that Eq.~(\ref{rep1}) can be understood 
classically as being due to 
a charge in uniform motion in vacuum,\cite{fermi40}
Eq.~(\ref{rep2}) can be understood as due to a charge is in uniform 
motion in a medium. Because the motion is uniform, the time dependence
can be eliminated in favor of a spacial derivative opposite to 
the direction of motion and multiplied by the speed of the particle. 
For motion in the $z$-direction
\begin{equation}
\frac{\partial}{\partial t}\to -v_0\frac{\partial }{\partial z}
\end{equation}

Therefore, if we consider the electric field
\begin{equation}
{\bm E}=-\nabla\phi-\frac{1}{c}\frac{\partial {\bm A}}{\partial t}
\to
-\nabla\phi+\frac{v_0}{c}\frac{\partial {\bm A}}{\partial z},
\end{equation}
Eq.~(\ref{rep1}) follows from the substitution $A_z=({v_0}/{c})\phi$,
whereas Eq.~(\ref{rep2}) follows by making the correct substitution in the
presence of a medium
\begin{equation}
{A_z}=\frac{v_0}{c}\epsilon\phi\;,
\end{equation}
which in Fourier space gives 
%$$
%{\bm E}=-\nabla\phi+\hat z \frac{v_0^2}{c^2}\frac{\partial {\epsilon\phi}}{\partial z}\;.
%$$
%Or, in Fourier space,
\begin{eqnarray}\label{Eq:class}
{\bm E}({\bm q},\omega)
&=&
\left[
-i{\bm q}+\hat z \frac{\epsilon(\omega) v_0^2}{c^2}iq_z
\right]\phi({\bm q},\omega)\;,
%\nonumber 
%\\
%&=&
%\left[
%-i{\bm q_\perp}-i\hat {\bm z} q_z\left(1-\epsilon(\omega)\frac{v_0^2}{c^2}\right)
%\right]\phi({\bm q},\omega),
\end{eqnarray}
which is equivalent to Eq.~(\ref{rep2}).

Because Eq.~(\ref{rep2}) depends on the macroscopic dielectric 
function the ratio $\theta_M/\theta_E$, which formerly was a function
only of $v_0$, will now show material dependence. This is seen
from the generalization of Eqs.~(\ref{Eq:F}) and (\ref{Eq:G}), the
equality of which gives the magic angle. Instead of 
Eq.~(\ref{Eq:F}) for $F(\alpha_c)$ we
now have
\begin{equation}\label{Eq:FGLG}
F(\alpha_c)\equiv\int_0^{\alpha_c}d\theta \theta 
\frac{\theta^2}{{|\theta^2+\theta_E^2g|}^2}\;,
\end{equation} 
and, instead of Eq.~(\ref{Eq:G}) we now have
\begin{equation}\label{Eq:GGLG}
G(\alpha_c)\equiv2\theta_E^2{|g|}^2\int_0^{\alpha_c}d\theta \theta
\frac{1}{{|\theta^2+\theta_E^2g|}^2}\;,
\end{equation}
where
\begin{equation}
g=1-\epsilon(\omega) v_0^2/c^2
\end{equation}
is a complex number which replaces $1/\gamma^2$ in the vacuum
relativistic theory.

If one can calculate the macroscopic dielectric function of the sample
by some means~\cite{prangeetal05} then the material dependent magic angle
can be determined theoretically and compared to experiment. 
Furthermore, the correction to the magic angle given by the introduction 
of the macroscopic dielectric constant relative to the relativistic 
macroscopic ``vacuum value" of Jouffrey {\it et al}.~is seen to be typically 
positive 
(since ${\rm Re}[\epsilon]\lesssim 1$ and $0\lesssim {\rm Im}\, [\epsilon]$),
in rough agreement with observation.\cite{daniels03} In fact, it turns out
that the correction is always positive for the materials we consider and
is substantial only for low energy-loss where the dielectric function
differs substantially from its vacuum value. 
For modern EELS experiments which use relativistic
microscope energies and examine low energy-loss regions, the effect
of the dielectric correction on the magic angle should be large.

\begin{figure}
%\begin{center}
%\includegraphics[scale=0.35,angle=270]{Magic_BN.4a.ps}
%\includegraphics[scale=0.35,angle=270]{Magic_BN.4a.bw.ps}
\includegraphics[scale=0.35,angle=270]{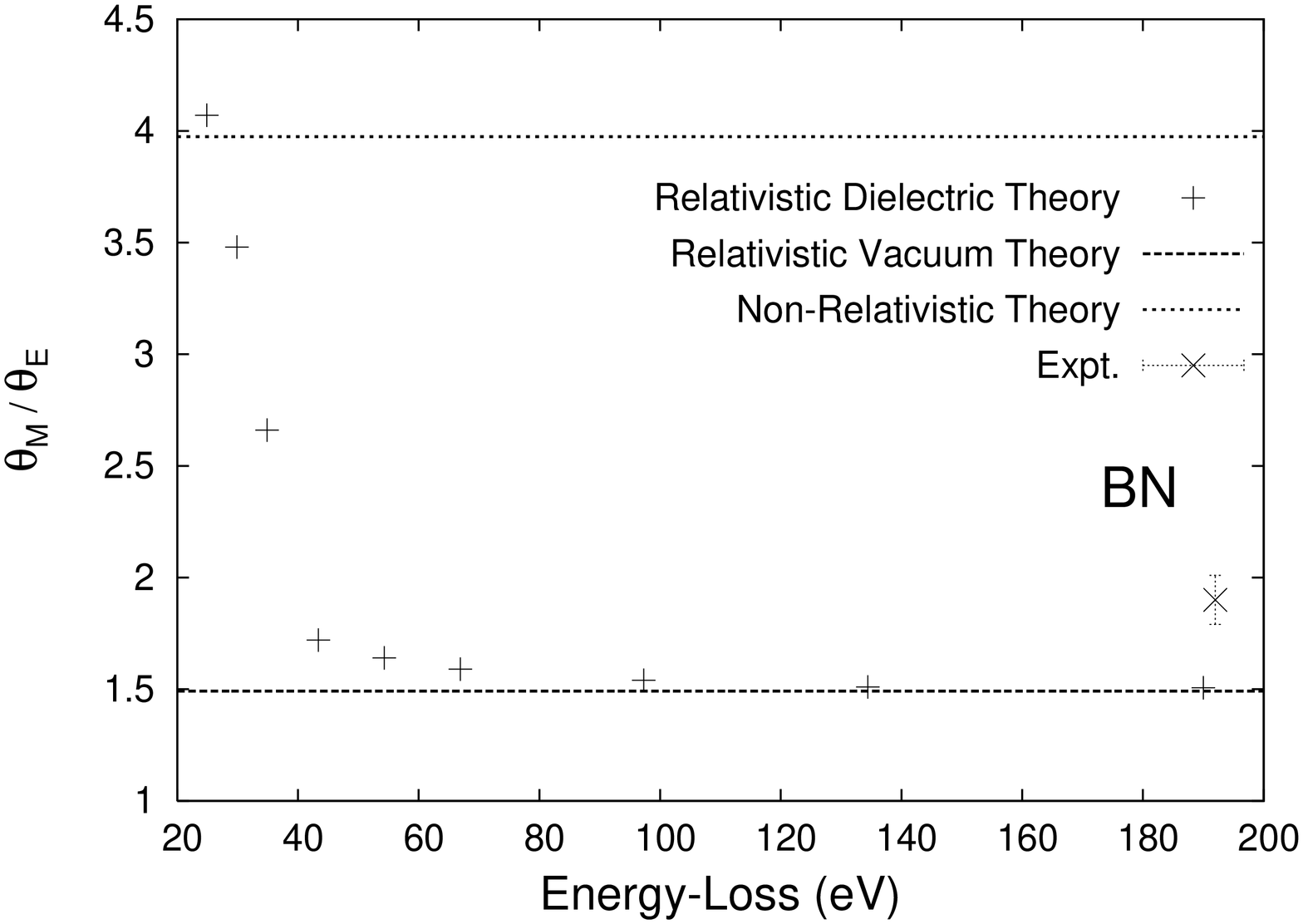}
\includegraphics[scale=0.35,angle=270]{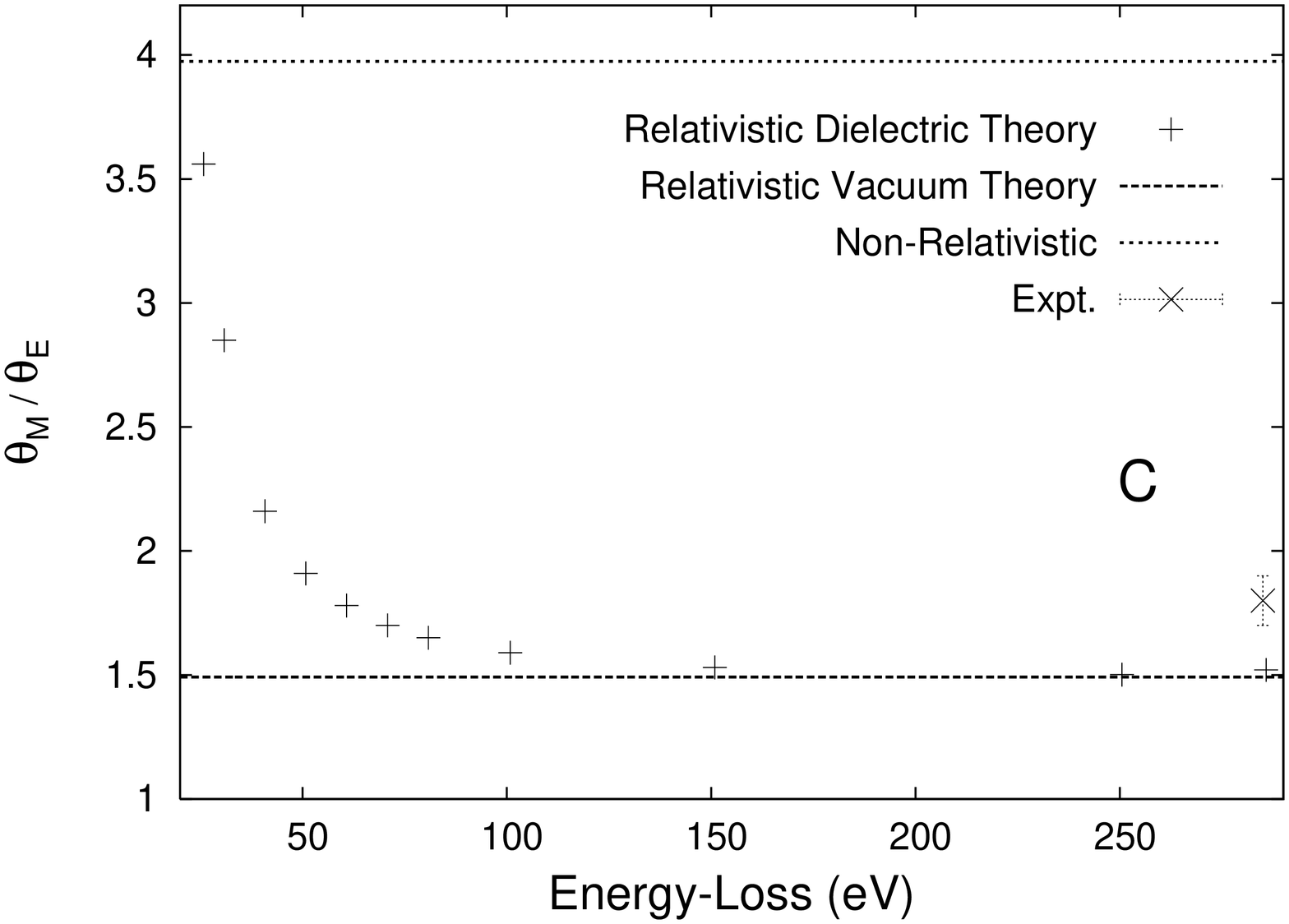}
\caption{\label{Fig:BN-C}The magic angle to characteristic angle 
ratio $\theta_M/\theta_E$ is compared for three differing theories
and one experiment.\cite{daniels03}
The materials considered in the figure are boron nitride (top figure)
and graphite (bottom). The microscope
voltage is fixed at 195 keV. Both the 
non-relativistic and relativistic vacuum theories show no dependence 
on the energy-loss and no dependence on the material. The
relativistic dielectric theory shows that the magic angle should deviate 
from the vacuum value by a significant amount in regions where 
the macroscopic dielectric response is substantial.}
%\end{center}
\end{figure}

Example calculations using our relativistic dielectric theory compared to both 
the relativistic vacuum theory of Schattschneider {\it et al}.~and to the 
non-relativistic vacuum theory are shown in 
Fig.~(\ref{Fig:BN-C}) 
%and (\ref{Fig:C}) 
for the materials boron nitride and graphite. The data of 
Daniels et {\it al}.~\cite{daniels03} is also shown in the figures. 
We have not attempted to estimate the true error bars for the data; the
error bars in the figure indicate only the error resulting from 
the unspecified finite convergence angle.

%\begin{figure}
%%\begin{center}
%%\includegraphics[scale=0.35,angle=270]{Magic.C.ps}
%%\includegraphics[scale=0.35]{diagram.eps}
%\caption{\label{Fig:C}The magic angle to characteristic angle 
%ratio $\theta_M/\theta_E$ is compared for three differing thoeries. 
%The material considered in the figure is graphite. The microscope
%voltage is fixed at 195 keV. Both the 
%non-relativistic and relativistic vacuum theories show trivial dependence 
%on the energy-loss and trivial dependence on the material. The
%relativistic dielectric theory shows that the magic angle should deviate 
%from the vacuum value by a significant amount in regions where 
%the macroscopic dielectric response is substantial.}
%%\end{center}
%\end{figure}
%
\section{Conclusions}

We have developed a fully relativistic theory of the magic angle
in electron energy loss spectra starting from the QED Hamiltonian of
the many body system. As with the single-particle theory of 
Jouffrey {\it et al.}~and 
Schattschneider {\it et al.}~we find a factor of two
``transverse" correction to the non-relativistic ratio 
$\theta_M/\theta_E$.
%which is relevant to EELS experiments
%starting from the QED Hamiltionan. 
We have also shown how 
macroscopic electrodynamic effects
can be incorporated into the relativistic
single-particle formalism of Schattschneider {\it et al}. In particular
we predict that these dielectric effects can be important for
determining the correct material-dependent magic angle
at low energy-loss, where the difference between the dielectric function
relative to its vacuum value is observed to be substantial.

Several other factors may be important for correctly describing the
energy loss dependence of the magic angle in anisotropic materials. In 
particular, we believe that further study of the many body effects 
(beyond a simple macroscopic dielectric model) via explicit calculations of the 
{\em microscopic} dielectric function 
and including time-dependant density functional/Bethe-Salpeter theory 
TDDFT/BSE\cite{ankudinov05} are an important next step in the description of
all EELS phenomena, including the magic angle.

%tend to increase the magic angle and
%introduce a sample dependence as observed experimentally.

\section{Acknowledgments}

We wish to thank K. Jorrisen for helpful comments and encouragement. 
This work is supported by National Institute of Standards and 
Technology (NIST) Grant 70 NAMB 2H003 (APS),
Department of Energy (DOE) Grant DE-FG03-97ER45623 (JJR), and was
facilitated by the DOE Computational Materials Science Network.

%\bibliography{magicbib}
%\bibliographystyle{apsrev}

\appendix*

\section{Relativistic Effects}
%\section{Appendix: Relativistic and Many-Body Effects}
%
% in this section I actually should be more careful with the 
% spin parts of the probe eelctrons (i.e. the spin dependence of the
% u(k) spinors), but the spin effects are, i believe, order z\alpha squared
% already which I am already ignoreing... Actually, I'm not positive about
% this. The spin-orbit should be order Z\alpha squared compared to the 
% rest... I should clean this up a bit maybe. but for now I like it because
% it shows how to make contact with the non-rel theory... also the issue
% of the potential seen by the probe... if I keep around all the electrons
% of the solid explictly the diffraction is off of nuclei and is elastic, etc.
% but when I go to the single particle picture of the solid, even tho
% I keep around the coulomb interaction between sample electron and probe 
% electron the potential "seen" by the probe is certainly modified and
% tho it is not exactly the same as that seen by the sample electron it
% maybe be only very very negligibly different...
%

Starting from Eq.~(\ref{hammy}) we write (the notation $H_0$ in the 
Appendix differs from that in the main body of the text):
\begin{eqnarray}\label{Eq:mny}
H_0&=&c{\bm \alpha}\cdot{\bm p}+{\bm \beta} mc^2 
+ 
v_{\textrm{e-core}}({\bm r})
\nonumber
\\
&+&
\sum_{i=1}^N
\left[
c{\bm \alpha^{(i)}} \cdot{\bm p}^{(i)}+{\bf \beta}^{(i)}mc^2
+
v_{\textrm{e-core}}({\bm r}^{(i)})
\right]
\nonumber
\\
&+&
\frac{1}{2}\sum_{1=i\neq j=1}^N\frac{e^2}{|{\bm r}^{(i)}-{\bm r}^{(j)}|}
+
H_\textrm{rad},
\end{eqnarray}
and
\begin{equation}
U=
e{\bm \alpha\cdot{\bm A}({\bm r})
+
\sum_{i=1}^N
\frac{e^2}{|{\bm r}-{\bm r^{(i)}}|}
+
\sum_{i=1}^N e{\bm \alpha^{(i)} }\cdot{\bm A}({\bm r^{i}}} )
\;.
\end{equation}
We are interested in matrix elements of the perturbation
\begin{equation}
U+UG_0U+\ldots
\end{equation}
between eigenstates of the unperturbed Hamiltonian 
\begin{equation}
|I\rangle = \ket{k_i}\ket{\Psi_i}\ket{0} \quad {\rm and}\quad
|F\rangle = \ket{k_f} \ket{\Psi_f} \ket{0}\;.
\end{equation}
The difference between the many-body case and the 
single-particle theory of the sample is that the wavefunction
of the sample now depends on $N$ electron coordinates, instead of one
effective coordinate. Also we see that the only 
potential ``seen" by the probe (i.e., included in the unperturbed
probe Hamiltonian) is the $v_{\textrm{e-core}}$ potential. 
This is to be contrasted
with the ``unperturbed" sample Hamiltonian which includes not only the
$v_{\textrm{e-core}}$ but also the Coulomb interactions between all the 
sample electrons.

Consequently, working with a unit volume and 
proceeding exactly as in the single-particle case, we find
a ``longitudinal" contribution to the matrix element
\begin{equation}\label{Eq:LMB}
M_L=
%\frac{4\pi e^2}{Vq^2}
\frac{4\pi e^2}{q^2}
u^{\dagger}({\bm k_f})u({\bm k_i}) \bra{\Psi_f}
\sum_{i=1}^N
e^{i{\bm q}\cdot{\bm r^{(i)}}}
\ket{\Psi_i},
\end{equation}
and a ``transverse" contribution
\begin{equation}
%M_T=\frac{4\pi e^2}{V}
%\frac{1}{\omega^2/c^2-q^2}
M_T=
\frac{4\pi e^2}{\omega^2/c^2-q^2}
u^{\dagger}({\bm k_f}){\bm \alpha_T} u({\bm k_i})
\cdot
\bra{\Psi_f}
\sum_{i=1}^N{\bm \alpha^{(i)} } e^{i{\bm q}\cdot{\bm r^{(i)}}}
\ket{\Psi_i}\;,
\end{equation}
where
\begin{equation}
{\bm \alpha_T}={\bm \alpha} -{\bm q}
\frac{ {\bm q}\cdot{\bm \alpha} } {q^2}\;,
\end{equation}
and where the $u({\bm p})$ are the usual free-particle Dirac
spinors, normalized such that
\begin{equation}
u^\dagger({\bm p}) u({\bm p}) = 1\;.
\end{equation}

The two matrix elements $M_L$ and $M_T$ are to be summed and then
squared, but before proceeding with this plan we make the following 
useful definitions:
The transverse Kronecker delta function (transverse to momentum-transfer)
\begin{equation}
\delta_T^{ij}=\delta^{ij}-\frac{q^iq^j}{q^2}\;;
\end{equation}
the (Fourier transformed) density operator
\begin{equation}
n({\bm q})=\sum_{i}^Ne^{-i{\bm q}\cdot{\bm r^{(i)}}}\;;
\end{equation}
and the (Fourier transformed) current operator
\begin{equation}
{\bm j}({\bm q})=\sum_{i}^N c{\bm \alpha^{(i)}}e^{-i{\bm q}\cdot{\bm r^{(i)}}}\;.
\end{equation}

Next, we recall some properies of the Dirac spinors $u({\bm p})$ and of
Dirac matrices which we will presently find useful:
\noindent
i) There are four independent spinors 
$\left\{u^{(1)},u^{(2)},u^{(3)},u^{(4)}\right\}$,
the first two of which will refer to positive energy solutions, and
the second two of which will refer to negative energy solutions (and are
not used in this calculation);

\noindent
ii) the positive energy spinors satisfy a ``spin sum" 
\begin{eqnarray}
\sum_{s=1}^2u^{(s)}({\bm p}){u^{(s)}}^\dagger({\bm p})&=&
\frac{1}{2E(p)}
\left(
E(p)+c{\bm \alpha}\cdot{\bm p}+{\bm \beta}mc^2
\right)
\nonumber
\\
&\equiv&
\frac{1}{2E(p)}
\left(
E(p)+h_D({\bm p})
\right)\;,
\end{eqnarray}
where $E(p)=\sqrt{p^2c^2+m^2c^4}$;
%and the negative energy spinors satisfy
%$$
%\sum_{s=3,4}u^{(s)}(k){u^{(s)}}^\dagger(k)=\frac{1}{2|E(k)|}
%\left(
%|E(k)|-c{\bm \alpha}\cdot{\bm k}-{\bm \beta}mc^2
%\right)
%$$
%and, of course
%$$
%\sum_{s=1,4}u^{(s)}(k){u^{(s)}}^\dagger(k)=1\;;
%$$

\noindent
iii) the Dirac matrixes satisfy the trace identities
\begin{eqnarray}
{\rm Tr}({\bm \alpha_i}{\bm \alpha_j})&=&4\delta_{ij}, \\
%\end{equation}
%\begin{equation}
{\rm Tr}({\bm \alpha_i}{\bm \alpha_j}{\bm \alpha_k}{\bm \alpha_l})&=&
4\left(
\delta_{ij}\delta_{kl}-\delta_{ik}\delta_{jl}+\delta_{il}\delta_{jk}
\right), \\
%\end{equation}
%\begin{equation}
{\rm Tr}({\bm \alpha_T^i}{\bm \alpha^j})
&=&
{\rm Tr}({\bm \alpha_T^i}{\bm \alpha_T^j})
=4\delta_T^{ij}, \\
%\end{equation}
%\begin{equation}
{\rm Tr}({\bm \alpha^i}{\bm \alpha^j_T}{\bm \alpha^k}{\bm \alpha^l_T})&=&
4\left( \delta^{ij}_T\delta^{kl}_T-\delta^{ik}\delta^{jl}_T
+\delta^{il}_T\delta^{jk}_T
\right);\ \ 
\end{eqnarray}

\noindent
iv) Finally, we note that in this calculation there are many simplifications
due to the fact that $\omega<<mc^2<E({\bm k_I})\approx E({\bm k_F})$. For
example,
\begin{eqnarray}
\frac{1}{2E_iE_j}(E_iE_j+m^2c^4+c^2{\bm k_i}\cdot{\bm k_j})
&=&
\nonumber
\\
1-\frac{\omega}{E(k_i)}+O({\frac{\omega^2}{E(k_i)^2}})
\approx 1\;;
\nonumber
\end{eqnarray}
throughout the calculation we ignore terms of order $\omega/E({\bm p_I})$
%
%$$
%{\rm Tr}({\bm \alpha^i}{\bm \alpha_T^j}{\bm \alpha^k}{\bm \alpha_T^l})
%=
%4(\delta_{T}^{ij}\delta_{T}^{kl}-\delta^{ik}\delta_{T}^{jl}
%+\delta_{T}^{il}\delta_{T}^{jk})
%$$
%where we employ a convienient shorthand $\delta_T^{ij}=\delta^{ij}-q^iq^j/q^2$.
%Whew.
Using these identities it is easy to see that 
\begin{equation}
\frac{1}{2}\sum_{s_i=1}^2\sum_{s_f=1}^2{|M_L|}^2
%={\left(\frac{4\pi e^2}{Vq^2}\right)}^2{|\bra{\Psi_F}n_q^\dagger\ket{\Psi_I}|}^2
={\left(\frac{4\pi e^2}{q^2}\right)}^2{|\bra{\Psi_F}n_q^\dagger\ket{\Psi_I}|}^2
\end{equation}
which has the same form as in the non-relativistic case 
(up to order $\omega/E_i$); the squared  
matrix element is much simplified by the sum over final probe-spin and
average over initial probe-spin. Of course, the matrix element itself
is completely general in terms of probe-spin, 
but many simplification arise from ignoring the probe-spin and
exploiting the spin-sums.

Continuing on to the transverse matrix element--and including a
few more of the details ($a$, $b$, $c$, and $d$ are Dirac indices)--we find
\begin{eqnarray}
&&
\frac{1}{2}\sum_{s_i=1}^2\sum_{s_f=1}^2{|M_T|}^2
=
\frac{1}{2}\sum_{s_i=1}^2\sum_{s_f=1}^2
%{\left(\frac{4\pi e^2}{V(\omega^2/c^2-q^2)}\right)}^2\times
{\left(\frac{4\pi e^2}{\omega^2/c^2-q^2}\right)}^2
\nonumber
\\
&&
\times
{u({\bm k_f})^{(s_f)}_a}^*{\bm \alpha_T^m}^{ab}u({\bm k_i})^{(s_i)}_b
{u({\bm k_i})^{(s_i)}_c}^*{\bm \alpha_T^n}^{cd}u({\bm k_f})^{(s_f)}_d
\nonumber
\\
&&\times
\bra{\Psi_F}{j_m({\bm q})}^\dagger\ket{\Psi_I}
\bra{\Psi_I}j_n({\bm q})\ket{\Psi_F}
\nonumber
\end{eqnarray}
\begin{eqnarray}
&&=
%{\left(\frac{4\pi e^2}{V(\omega^2/c^2-q^2)}\right)}^2
{\left(\frac{4\pi e^2}{\omega^2/c^2-q^2}\right)}^2
\bra{\Psi_F}{j_m({\bm q})}^\dagger\ket{\Psi_I}
\bra{\Psi_I}j_n({\bm q})\ket{\Psi_F}
\nonumber
\\
&&\times
{\rm Tr}\left(
(E(k_f)+h_D({\bm k_f}))
{\bm \alpha_T^m}
(E(k_i)+h_D({\bm k_i}))
{\bm \alpha_T^n}
\right)
\nonumber
\\
%&&={\left(\frac{4\pi e^2}{V(\omega^2/c^2-q^2)}\right)}^2
&&={\left(\frac{4\pi e^2}{\omega^2/c^2-q^2}\right)}^2
{\left|
\bra{\Psi_I}\frac{{\bm v_T}\cdot{\bm j({\bm q})}}{c^2}\ket{\Psi_F}
\right|}^2\;.
\end{eqnarray}

For the cross term we find
%the cross-term
%At this point--we might even be able to guess the result--it is no surprise
%to find for the cross-term
\begin{eqnarray}
&&
\frac{1}{2}\sum_{s_i=1}^2\sum_{s_f=1}^2M_L M_T^*
= {\left({4\pi e^2}\right)}^2\frac{1}{q^2(\omega^2/c^2-q^2)}\nonumber \\
&&
%{\left(\frac{4\pi e^2}{V}\right)}^2\frac{1}{q^2(\omega^2/c^2-q^2)}
\times \bra{\Psi_F}n^{\dagger}({\bm q})\ket{\Psi_I}
\bra{\Psi_I}\frac{{\bm j}({\bm q})\cdot{\bm v_T}}{c^2}\ket{\Psi_F}
\;.
\end{eqnarray}
Thus
%all in all,
we have finally derived an expression for the
relativistic many-body summed-then-squared matrix elements summed
and averaged over spins,
\begin{eqnarray}\label{Mfull}
&&
\frac{1}{2}\sum_{s_i}\sum_{s_f}{|M_L+M_T|}^2=
{\left(\frac{4\pi e^2}{V}\right)}^2
\nonumber
\\
%&&
%{\left(\frac{4\pi e^2}{V}\right)}^2
%\left\{
%\frac{1}{q^4}{|{\bra{\Psi_I}n({\bm q})\ket{\Psi_F}}|}^2
%+\right.
%\nonumber
%\\
%&&
%\frac{1}{(\omega^2/c^2-q^2)^2}{|\bra{\Psi_I}\frac{{\bm v_T}\cdot{\bm j}({\bm q})}{c^2}\ket{\Psi_F}|}^2
%+
%\nonumber \\
%&&\left.
%\frac{1}{q^2(\omega/c^2-q^2)^2}2{\rm Re}[\bra{\Psi_F}n^\dagger({\bm q})
%\ket{\Psi_I}\right. \nonumber \\
%&&
%\left.\bra{\Psi_I}\frac{{\bm v_T}\cdot{\bm j({\bm q})}}{c^2}\ket{\Psi_F}]\right\}
%\nonumber
%\\
&&
%={\left(\frac{4\pi e^2}{V}\right)}^2 
\times{\left|
\frac{\bra{\Psi_I}n({\bm q})\ket{\Psi_F}}{q^2}
+\frac{\bra{\Psi_I}{\bm v_T}\cdot{\bm j}({\bm q})\ket{\Psi_F}}{\omega^2-q^2c^2}
\right|}^2
\\
&&
={\left[\frac{4 \pi e^2}{V(\omega^2/c^2-q^2)}\right]}^2
{\left|
\bra{\Psi_I}
n({\bm q})-\frac{{\bm v_0}\cdot{\bm j}({\bm q})}{c^2}
\ket{\Psi_F}
\right|}^2\;.
\nonumber
\end{eqnarray}
The last equality follows from
\begin{equation}
{\bm q}\cdot\bra{\Psi_I}{\bm j}({\bm q})\ket{\Psi_F}
=\omega\bra{\Psi_I}n({\bm q})\ket{\Psi_F},
\end{equation}
which itself follows by considering the commutator analogous
to that of Eq.~(\ref{commie}).

The final line of Eq.~(\ref{Mfull}) is quite pleasing since we have 
found that if we can ``ignore" the spin of the probe particle, we may 
as well have started by taking matrix elements between electronic
states only 
%[the photons have dropped out entirely!]
of the much simpler interaction Hamiltonian
\begin{equation}\label{pertq}
U'=\int d^3x \left[
n({\bm x})\phi_\omega({\bm x}-{\bm x_p})-
\frac{{\bm j}({\bm q})\cdot{\bm A_\omega}({\bm x}-{\bm x_p})}{c^2}
\right]\;,
\end{equation}
where the fields $\left\{\phi_\omega,{\bm A_\omega}\right\}$ are just the
$e^{-i\omega t}$ components of the {\em classical} 
field of a point charge of velocity 
${\bm v_0}$ in the Lorentz gauge, and where 
\begin{equation}\label{xden}
n({\bm x})=\sum_i^N\delta({\bm x}-{\bm x^{(i)}})\;,
\end{equation}
and
\begin{equation}\label{xcur}
{\bm j}({\bm x})=\sum_i^N c{\bm \alpha^{(i)}}
\delta({\bm x}-{\bm x^{(i)}})\;.
\end{equation}
That is, if we take Eq.~(\ref{pertq}) as our starting point and proceed 
in the usual way, we will find that our squared matrix elements are exactly 
the same as what we know to be correct from Eq.~(\ref{Mfull}).
The photons have dropped out entirely!
\vfill
\end{document}